\documentclass[preprint,aps,showpacs,onecolumn]{revtex4} %,floatfix
\usepackage{amsmath}
\usepackage{amsfonts}
\usepackage{amssymb}
\usepackage{bm}
\usepackage[dvips]{epsfig}
\usepackage[dvips]{graphicx}
\usepackage{float}
\usepackage{dcolumn}
\usepackage{ifpdf}
\usepackage{dcolumn}
\usepackage{multirow}

\newcommand{\be}{\begin{equation}}
\newcommand{\ee}{\end{equation}}
\newcommand{\bes}{\begin{subequations}}
\newcommand{\ees}{\end{subequations}}
\newcommand{\ben}{\begin{eqnarray}}
\newcommand{\een}{\end{eqnarray}}

\begin{document}
\title{Suppression of two-bounce windows in kink-antikink collisions}
 \author{F. C. Simas$^{1}$, Adalto R. Gomes$^{2}$, K. Z. Nobrega$^{3}$, J. C. R. E. Oliveira$^{4,5}$}
 \email{simasfc@gmail.com,argomes.ufma@gmail.com,bzuza1@yahoo.com.br,joanaespain@gmail.com}
 \noaffiliation
\affiliation{
$^1$ Centro de Ci\^encias Agr\'arias e Ambientais-CCAA, Universidade Federal do Maranh\~ao (UFMA), 65500-000, Chapadinha, Maranh\~ao,
Brazil\\
$^2$ Departamento de F\'\i sica, Universidade Federal do Maranh\~ao (UFMA) \\
Campus Universit\'ario do Bacanga, 65085-580, S\~ao Lu\'\i s, Maranh\~ao, Brazil\\
$^3$ Departamento de Eletro-Eletr\^onica, Instituto Federal de Educa\c c\~ao, Ci\^encia e
Tecnologia do Maranh\~ao (IFMA), Campus Monte Castelo, 65030-005, S\~ao Lu\'is, Maranh\~ao, Brazil \\
$^4$ Centro de F\'\i sica do Porto, Rua do Campo Alegre 687, 4169-007 Porto, Portugal\\
$^5$Departamento de Engenharia F\'\i sica da Faculdade de Engenharia da Universidade do Porto,
 Rua Dr. Roberto Frias, s/n, 4200-465 Porto, Portugal}
\noaffiliation

\begin{abstract}
We consider a class of topological defects in $(1,1)$-dimensions with a deformed $\phi^4$ kink structure whose stability analysis leads to a Schr\"odinger-like equation with a zero-mode and at least one vibrational (shape) mode. We are interested in the dynamics of kink-antikink collisions, focusing on the structure of two-bounce windows. For small deformation and for one or two vibrational modes, the observed two-bounce windows are explained by the standard mechanism of a resonant effect between the first vibrational and the translational modes. With the increasing of the deformation, the effect of the appearance of more than one vibrational mode is the gradual disappearance of the initial two-bounce windows. The total suppression of two-bounce windows even with the presence of a vibrational mode offers a counterexample from what expected from the standard mechanism. For even larger deformation, some two-bounce windows reappear, but with a non-standard structure.
\end{abstract}

\pacs{ 11.10.Lm, 11.27.+d, 98.80.Cq}

\maketitle

%%%%%%%%%%%%%%%%%%%%%%%%%%%%%%%%%%%

\section{ Introduction } Solitary waves are important objects of investigation in several areas of nonlinear physics in all scales, from low-energy \cite{dauxois} to high-energy physics \cite{top}. The simplest solitary wave solution obtained with scalar fields is the $(1,1)$ dimensional kink. The embedding of a kinklike defect in three spatial dimensions gives rise to a domain wall, a topological defect separating a region of space in two volume domains. The initial physical conditions originating domain walls often favor the emergence of multiple domains separated by dynamical wall networks. Deviations from standard domain wall models, allowing walls with different energy or topology or biased vacuum values modify considerably the network structure, originating different wall patterns \cite{mod}.  A first-order phase transition in the early universe could generate bubbles of the broken-symmetry phase. The study of collapse of collapsing domain bubble \cite{zko} contributed for the discovery of breather solutions. In the regime of high bubble nucleation rate, one can consider the collision of two bubbles as in flat spacetime with $SO(2,1)$ symmetry \cite{vac}.  For very large bubbles, the collision process can include in addition planar symmetry, an ingredient also used in the context of branes \cite{brane}. In a Minkowski background, this reduces the background dynamics of colliding domain walls to that of a $K\bar K$ pair in $(1,1)$ dimensions, as used for instance in the study of the effects of small initial quantum fluctuations in nucleated bubbles in collision \cite{bubble}.

Kink and antikink solutions can be obtained for instance in the renormalizable and non-integrable $\phi^4$ theory.  Despite its simplicity, in this theory and in several other nonintegrable models the process of $K\bar K$ collisions can be surprisingly rich, when analyzed as a function of the initial velocity of approximation \cite{k,s,bk,csw,cps,cp,aom,gh,m,gk,w1,gll}. For large initial velocity the pair $K\bar K$ recedes from each other whereas for small initial velocity a $K\bar K$ bion state \cite{a} is formed. For intermediate velocities, however, the richness of the collision is revealed with windows in velocity (called bounce windows) where an integer number of bounces do appear before the components of the $K\bar K$ pair recede from each other. If one zooms in the border of a region of a certain number of bounces, a new window shows up with a higher number of bounces in a kind of fractal structure \cite{aom}. The present work deals with the simplest effect of two-bounce windows. Two-bounce windows were also observed in collisions between kinks and defects \cite{fkv,fkv2,fkv3,gh2} and in collisions of orthogonally polarized vector solitons in birefringent optical fibers \cite{ty1,ty2,gh3}. For a good review for effects of nonlinearity in classical field theory for nonintegrable systems, see \cite{bk2} and references therein.

According to Campbell, Sch\" onfeld and Wingate (CSW) \cite{csw}, a resonance effect is the mechanism behind the appearance of two-bounce windows. There the separation of the formed pair $K\bar K$ after the second bouncing is due to the change from the first vibrational mode and consequent restoration to the translational mode in a resonant mechanism.  A counter-example of this mechanism was found for the $\phi^6$ model and presented in Ref. \cite{phi6}, where it was shown that two-bounce windows could be obtained even in the absence of a vibrational mode, but as a result of collective mode produced by the pair $K\bar K$. For more results with this model, see Ref. \cite{gkl}.

In this paper we present another counter-example of the CSW mechanism, in which two-bounce windows disappear completely despite the presence of vibrational modes.
 For this, in section II we consider a class of deformed kinks \cite{dutra} driven by a parameter $b$ that recover the $\phi^4$ kink (for $b=0$) and in the limit $b\to\infty$ leads to a double kink \cite{bmm}. The model has always a zero mode and at least one vibrational mode. The number of vibrational modes grows with $b$. In section. III we consider the structure of $K\bar K$ scattering. The analysis shows the gradual disappearing of the two-bounce windows until their total suppression for a specific range of $b$ values. In section IV we present our main conclusions, including the connection between our finding, the appearance of an additional vibrational mode and the structure of the potential of linear perturbations.

\section {The model} We start with the action
\begin{eqnarray}
S = \int dxdt \biggl(\frac12 \partial_{\mu} \phi \partial^{\mu} \phi - V(\phi)\biggr),
\end{eqnarray}
where the $\phi$ is a real scalar field and $V(\phi)$ is the potential.
The equation of motion is $\phi_{tt}-\phi_{xx}+V_\phi=0$, where $V_\phi\equiv{dV}/{d\phi}$. The construction of static kinks $\phi_S$ with the first-order formalism requires the introduction of a superpotential $W(\phi)$. If the potential has the form $V(\phi)=1/2 W_\phi^2$, then the solutions of the first-order equation ${d\phi}/{dx}=\pm W_\phi$ are also solutions of the second order equation of motion.
The defects formed with this prescription minimize energy and are known as $BPS$ defects \cite{bps1,bps2}.
The $\phi^4$ model is the archetype of the construction of kink defects in non-integrable theories. It is characterized by a superpotential given by $W_\phi=1-\phi^2$,
with a solution given by $\phi(x) = \pm \tanh(x-x_0)$, where $x_0$ is the center of the kink. Stability analysis is a standard procedure and considers small fluctuations $\phi(x,t)=\phi_S+\eta(x)e^{i\omega t}$. This results in a Schr\"odinger-like equation
$-\eta_{xx}+V_{sch}(x)\eta=\omega^2\eta$ with $V_{sch}(x)=V_{\phi\phi}(\phi_S(x))$. With the introduction of the superpotential, it can be shown that the Hamiltonian is positive definite and tachyonic modes are absent.

\begin{figure}
\includegraphics[{angle=0,width=10cm,height=9cm}]{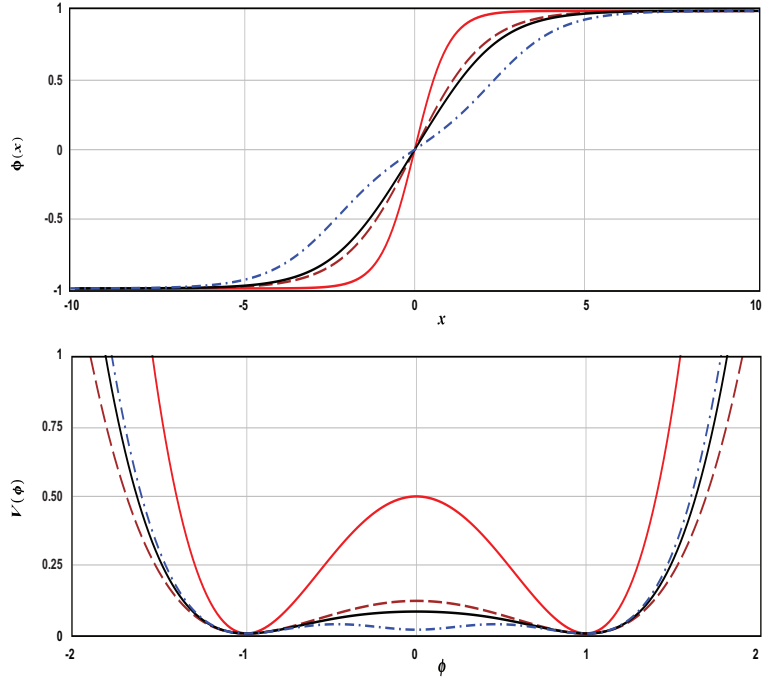}
\caption{ a) Field configurations $\phi(x)$  and b) Corresponding potential $V(\phi)$. The figures are for fixed $b=0$ (red solid), $b=1.05$ (brown dash), $b=1.5$ (black solid) and $b=5.0$ (blue dash-dotted).}
\label{sol_1}
\end{figure}

An interesting class of deformed kinks was considered in Ref. \cite{dutra}, where the properties of the defects where controlled by a parameter. The defect has the following scalar field profile
\begin{equation}
\phi(x)={\frac {\sinh \left( x \right) }{\cosh \left( x \right)+b}},
\label{solution}
\end{equation}
where an antikink solution may be obtained by the space reflection to get $\phi_{\bar{K}}=-\phi_K$. The dimensionless parameter $b$ regulates the appearance of a double kink character. Fig. \ref{sol_1}a depicts some plots of $\phi(x)$ for several values of $b$.
%%%%%%%%%%%%%%%%%%%%%%%%%%%%%%%%%%%%%%%%%%%%%%%%%%%%%%%%%%%%%%%%%%%%%
%%%%%%%%%%%%%%%%%%%%%%%%%%%%%%%%%%%%%%%%%%%%%%%%%%%%%%%%%%%%%%%%%%%%%
Such solution can be obtained by the superpotential \cite{dutra}
\be
W_\phi =  - \frac {   \left( {b}^{2}-1 \right) {\phi}^{2}+1-b
\sqrt {({b}^{2}-1){\phi}^{2}+1} }{ \left( {b}^{2}-1
 \right) }.
\ee
The Fig. $\ref{sol_1}b$ depicts the change in shape of the potential $V(\phi)$ corresponding to the several solutions shown in Fig. $\ref{sol_1}a$. Note that $b=0$ recovers the $\phi^4$ potential. We are interested in the region $b>1$.
%%%%%%%%%%%%%%%%%%%%%%%%%%%%%%%%%%%%%%%%%%%%%%%%%%%%%%%%%%%%%%%%%%%%%
\begin{figure}
\includegraphics[{angle=0,width=8cm}]{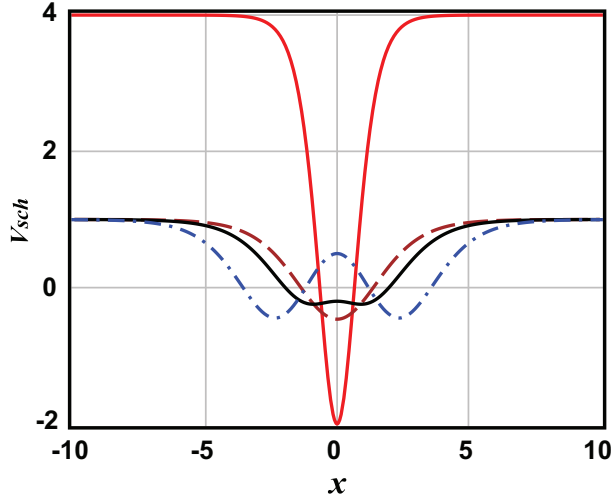}
\caption{Schr\"odinger-like potential $V_{sch}(x)$. The fixed values of $b$ and conventions are the same of Fig. \ref{sol_1}.}
\label{potential}
\end{figure}
%%%%%%%%%%%%%%%%%%%%%%%%%%%%%%%%%%%%%%%%%%%%%%%%%%%%%%%%%%%%%%%%%%%%%

Fig. \ref{potential} presents some plots of $V_{sch}(x)$ for the same parameters and conventions of Fig. \ref{sol_1}. Note from the figure that for all values of $b\geq 0$ there is a possibility of occurrence of bound states.  This potential for $b=0$ has two eigenvalues, corresponding to the translational ($\omega=0$) and vibrational ($\omega=\sqrt{3}$) modes. From the structure of the potential we note that these bound states must be centered at $x=0$, corresponding to the minimum of $V_{sch}$. This behavior of $V_{sch}$ is maintained for $b=1.01$, where a lower minimum at $x=0$ is accompanied by an increasing of the potential with $|x|$ which asymptotes to a lower maximum in comparison to the case $b=0$. From $b=1.01$ until $b=1.4$ the amplitude of the local minimum at $x=0$ is reduced. For $b\gtrsim 1.4$ the point $x=0$ turns to be a local maximum, with the appearance of two local minima in the potential. In this region, the larger is $b$, the larger is the local maximum at $x=0$, tending for $b\gg1$ to the formation of two trapping regions separated by a large potential barrier, the necessary conditions for the formation of a double kink.

We have solved the Schr\"odinger-like equation for several values of $b>1$. For $1<b\leq 1.1$ we have only one vibrational mode, as occurs to the $\phi^4$ model. The increasing of $b$ leads to the reduction of the first eigenvalue, tending to zero for large $b$. In this model we have an additional characteristic: the presence of two (for $1.2 \lesssim b \lesssim 4.1$) and three (for $b\gtrsim 4.2$) vibrational modes with the increasing of $b$. We noted also that the energy of the second vibrational mode decreases for large $b$.

%%%%%%%%%%%%%%%%%%%%%%%%%%%%%%%%%%%%%%%%%%%%%%%%%%%%%%%%%%%%%%%%%%%%%
\begin{figure}
\includegraphics[{angle=0,width=8cm}]{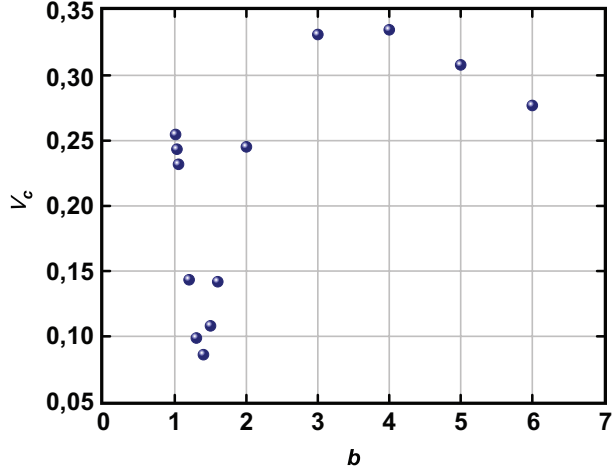}
\caption{Critical velocity $v_c$ versus b. }
\label{vc_b}
\end{figure}
%%%%%%%%%%%%%%%%%%%%%%%%%%%%%%%%%%%%%%%%%%%%%%%%%%%%%%%%%%%%%%%%%%%%%

\section { $K\bar K$ collisions  } We considered the collisions to be symmetric, with a deformed kink coming from $x\to -\infty$ with velocity $v$ and a deformed antikink  coming from $x\to +\infty$ with velocity $-v$. The initial conditions are
\begin{eqnarray}
\phi(x,0) &=& \phi_K(x+x_0,v,0)-\phi_K(x-x_0,-v,0)-1 \nonumber\\
\dot\phi(x,0) &=& \dot\phi_K(x+x_0,v,0)-\dot\phi_K(x-x_0,-v,0),
\end{eqnarray}
where $\phi_K(x+x_0,v,t)$ means a boost solution for the deformed kink.  We used a pseudospectral method on a grid with $2048$ nodes and periodic boundary conditions. We fixed $x_0=12$ as initial kink position, and we set the grid boundaries at $x_{max}=250$. For fixed value of $b$, we have the following possibilities: i) inelastic scattering (1-bounce) between the pair $K\bar K$ for $v>v_c$, defining the critical velocity $v_c$; ii) bion states for $v<v_c$, which are states where the pair $K\bar K$ remains trapped and irradiates continuously until being annihilated; iii) bounce windows, for $v\lesssim v_c$. Fig. \ref{vc_b} shows the plot of $v_c$ versus $b$. The plot shows that for $b>1$ the critical velocity decreases continuously with $b$ until achieving an absolute minimum around $b=1.4$. For $b>1.4$ there is an increasing of $v_c$ passing to a maximum, followed by a broad decreasing of $v_c$ for larger values of $b$. In this aspect, the behavior for $b>1.4$ is similar to the double sine-Gordon kink \cite{cps}.

The CSW mechanism is described by the relation \cite{csw}
\be\omega_1 T=2\pi m+\delta.
\label{Tm}\ee
 Here $T$ is the time between bounces and $\delta$ is a phase shift. The label $m$ characterizes the order of the two-bounce window, and is related to the number $M$ of oscillations of $\phi(x=0,t)$ between the two-bounces as $m=M-2$.

%%%%%%%%%%%%%%%%%%%%%%%%%%%%%%%%%%%%%%%%%%%%%%%%%%%%%%%%%%%%%%%%%%%%%%%
\begin{figure}[tbp]
\begin{center}
\includegraphics[{angle=0,width=10cm}]{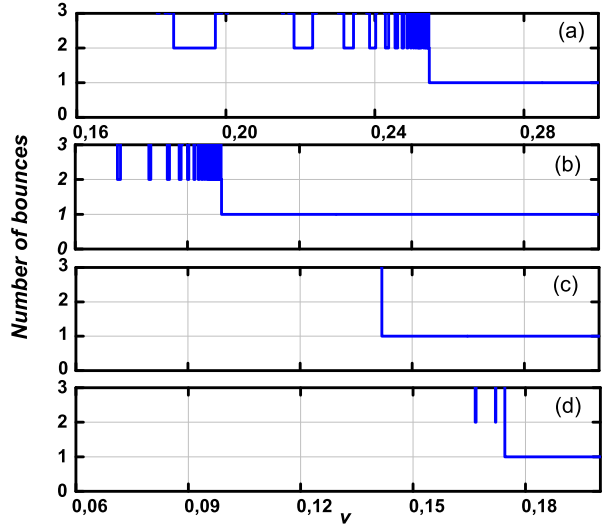}
\end{center}
\caption{Number of bounces versus initial velocity $v$ for a)  $b=1.01$, b) $b=1.3$, c)  $b=1.6$, d) $b=1.7$.}
\label{bouncv}
\end{figure}
%%%%%%%%%%%%%%%%%%%%%%%%%%%%%%%%%%%%%%%%%%%%%%%%%%%%%%%%%%%%%%%%%%%%%%

  Figs. \ref{bouncv}a-d show some results of the number of bounces versus initial velocity for fixed values of $b$.
   Fig. \ref{bouncv}a is for $b=1.01$, a case with just one vibrational mode, whereas Figs. \ref{bouncv}b-d corresponds to examples with two vibrational modes. The two-bounce windows observed in Fig. \ref{bouncv}a can be explained by the CSW mechanism. With the increasing of $b$, some of the two-bounce windows previewed are suppressed. We will show that one example of this corresponds to Fig. \ref{bouncv}b for $b=1.3$. With larger values of $b$ a region of total suppression of two-bounce windows is achieved (see Fig. \ref{bouncv}c for $b=1.6$). For even larger values of $b$,
some two-bounce windows are recovered (see Fig. \ref{bouncv}d).

{\tiny
%%%%%%%%%%%%%%%%%%%%%%%%%%%%%%%%%%%%%%%%%%%%%%%%%%%%%%%%%%%%%%%%%%%%%%%%%%%%%%%%%%%%%%%%%%%%%%%
\begin{table}[tbp]
\begin{tabular}{|c|c|c|c|c|}
\hline
$m$  & $\bar v$ & $\Delta v$  & $\beta$  & $\bar v^{pred}$ \\
\hline
1    & 0.1916  & 0.0111        & -       &  0.1966    \\ \hline
2    & 0.2208  & 0.004           & 2.84    &  0.2203    \\ \hline
3    & 0.2329  & 0.0025          & 2.92    &  0.2318    \\ \hline
4    & 0.2393  & 0.0015          & 2.89    &  0.2383    \\ \hline
\end{tabular}%
\caption{The center ($\bar v$), width ($\Delta v$), scaling ($\beta$) and predicted center ($\bar v^{pred}$) in the initial impact velocity for the first four observed windows in the two-bounces observed in collisions for $b=1.01$.}
\end{table}
%%%%%%%%%%%%%%%%%%%%%%%%%%%%%%%%%%%%%%%%%%%%%%%%%%%%%%%%%%%%%%%%%%%%%%%%%%%%%%%%%%%%%%%%%%%%%%%
}

%%%%%%%%%%%%%%%%%%%%%%%%%%%%%%%%%%%%%%%%%%%%%%%%%%%%%%%%%%%%%%%%%%%%%%%%%%%%%%%%%%%%%%%%%%%%%%%
\begin{table}[tbp]
\begin{tabular}{|c|c|c|c|c|}
\hline
$m$ &  $\bar v$   & $\Delta v$  & $\beta$  & $\bar v^{pred}$ \\
\hline
4   &  0.0716   & 0.0007              &  -      & 0.0796   \\ \hline
5   &  0.0799   & 0.0004           & 3.63    & 0.0843   \\ \hline
6   &  0.0848   & 0.0005          & 1.17    & 0.0875   \\ \hline
7   &  0.0880   & 0.0004          & 2.23    & 0.0897   \\ \hline
\end{tabular}%
\caption{The center ($\bar v$), width ($\Delta v$), scaling ($\beta$) and predicted center ($\bar v^{pred}$) in the initial impact velocity for the first four observed windows in the two-bounces observed in collisions for $b=1.3$.}
\end{table}
%%%%%%%%%%%%%%%%%%%%%%%%%%%%%%%%%%%%%%%%%%%%%%%%%%%%%%%%%%%%%%%%%%%%%%%%%%%%%%%%%%%%%%%%%%%%%%%

%%%%%%%%%%%%%%%%%%%%%%%%%%%%%%%%%%%%%%%%%%%%%%%%%%%%%%%%%%%%%%%%%%%%%%%
\begin{figure}[tbp]
\includegraphics[{angle=0,width=7cm,height=5cm}]{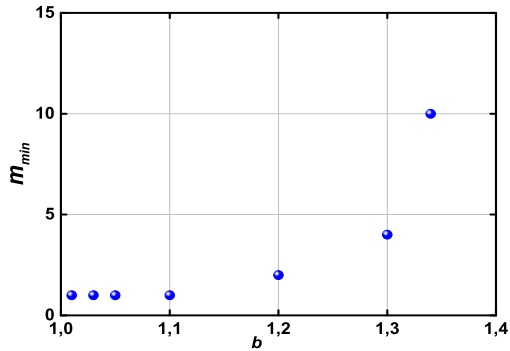}
\caption{Plot of the label $m_{min}$ of the first observed two-bounce windows versus $b$, showing  the gradual disappearing of the two-bounce windows until a total suppression around $b=1.35$.}
\label{mb}
\end{figure}
%%%%%%%%%%%%%%%%%%%%%%%%%%%%%%%%%%%%%%%%%%%%%%%%%%%%%%%%%%%%%%%%%%%%%%

\begin{figure}[tbp]
\begin{center}
\includegraphics[{angle=0,width=8cm}]{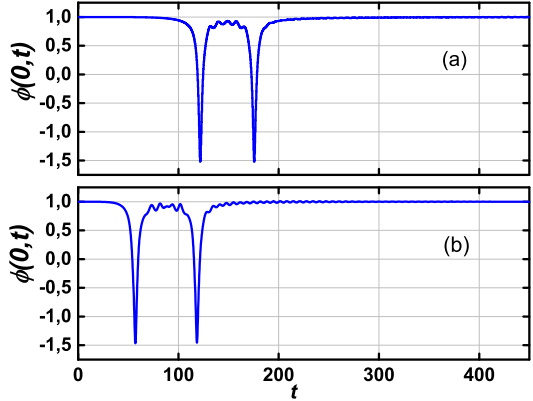}
\end{center}
\caption{Plots of $\phi(x=0,t)$ versus $t$, showing  a behavior characteristic a)  of the fourth two-bounce window ($m=4)$ for initial velocity $v=0.0713$ and $b=1.3$ (corresponding to a point from the first visible window in Fig. \ref{bouncv}b)); (b) of a two-bounce window, but with irregular oscillations, for $v=0.172$ and $b=1.7$ (corresponding to a point from the second visible window in Fig. \ref{bouncv}d).}
\label{wreg}
\end{figure}

  Tables I and II contains data of the first four observed two-bounce windows for $b=1.01$ and $b=1.3$, respectively (corresponding to Figs. \ref{bouncv}a-b).  Each window is characterized by a center $\bar v$ and a thickness $\Delta v$. From Table I we see that for $b=1.01$ the thickness of the two-bounce windows decrease with $m$, and that this agree with what observed in Fig. \ref{bouncv}b. Moreover for this value of $b$ we confirm the existence of a scaling relation $\Delta v\propto M^{-\beta}$ \cite{aom}, with $\beta\sim 2.89\pm0.05$. Now, Table II shows that for $b=1.3$,  two-bounce windows still appear (as already noted in Figs. \ref{bouncv}b), but their structure is affected by the presence of the extra vibrational state. Indeed, we verified that for $b=1.2$ the first two-bounce window is absent whereas Table II shows that for $b=1.3$ the same occurs with the first three ones. This shows that the suppression of the two-bounce starts in the lower value of $m$, labeled $m_{min}$, and that $m_{min}$ grows with $b$. This is confirmed by Fig. \ref{mb}. The figure shows that this region of suppression grows with $b$ and for $b\sim 1.35$ we observed total suppression of two-bounce windows.  The suppression is maintained for $1.35\lesssim b \lesssim1.6$. For $b > 1.6$ the two-bounce windows are partially recovered, but with the structure characterizing the two-bounce windows particularly distorted. This is illustrated in Figs. \ref{wreg}a-b. Note in Fig. \ref{wreg}a that $\phi(x=0,t)$ oscillates six times around the vacuum $\phi=1$, showing that this pattern corresponds to a two-bounce window labeled by $m=4$. This collision belongs  to the first two-bounce window on the left visible in Fig. \ref{bouncv}b, and some of its characteristics where described in the first line of Table II. On the other hand, Fig. \ref{wreg}b shows an irregular pattern of oscillation, and it is difficult to assign without ambiguity a value $m$ for this. This collision belongs to the second window on the left from Fig. \ref{bouncv}d.
  We also found that the scaling relation is affected with the appearance of an extra vibrational state. Indeed for $b=1.2$ we have $\beta\sim 3.1\pm 0.3 $ (showing an increasing in the uncertainty of $\beta$ when compared with the previous case) whereas Table II shows that for $b=1.3$ an agreement between the obtained values of $\beta$ is hardly possible, showing that there is no such scaling relation anymore.

In the region before suppression, we investigated some properties of the structure of the two-bounce windows, confirming the validity of  Eq. (\ref{Tm}) and the scaling relation \cite{csw} $T\propto (v^2_c-v^2)^{-1/2}$.
%%%%%%%%%%%%%%%%%%%%%%%%%%%%%%%%%%%%%%%%%%%%%%%%%%%%%%%%%%%%%%%%%%%%%%%
%%%%%%%%%%%%%%%%%%%%%%%%%%%%%%%%%%%%%%%%%%%%%%%%%%%%%%%%%%%%%%%%%%%%%%
We also used CSW theory to understand the measured center of two-bounce windows and to estimate values of lacked centers. Indeed, for each value of $b$,  it is possible to predict the centers as given by \cite{csw}
 \be
 v^2=v_c^2-\frac{r^2}{(\frac{2\pi}{\omega_1}m+\frac{\delta}{\omega_1})^2}.
 \ee
 Tables I to II show that the centers obtained numerically agree with those predicted by this method.

  \section { Conclusions  } Our analysis showed that the presence of at least one additional vibrational state concurs to spoil some of the resonant effect (according to the CSW mechanism) responsible for the formation of two-bounce windows. The effect of partial suppression of two-bounce windows was already previewed in the final considerations of Ref. \cite{csw} and observed for instance in the double sine-Gordon \cite{cps} and in $\phi^4$ kink-impurity interactions \cite{fkv}.  In the latter case a qualitative explanation of the effect could be made using collective coordinates \cite{col1,col2,col3,s}. Here, however, the quite intricate Schr\"odinger-like potential makes it impracticable to obtain explicit expressions for the vibrational eigenmodes, necessary for the application of the collective coordinates method. Nevertheless, a comparison between the qualitative results of both systems shows  that indeed the extra vibrational states are the ones responsible for the suppression of the two-bounce windows in the present case.
   A remarkable property of  this model, the effect of total suppression of two-bounce windows,
  was observed in a continuous range of $b$ (for $1.35\lesssim b \lesssim1.6)$ strictly connected to the beginning of changing of shape in the Schr\"odinger potential. Indeed, for $b \sim 1.4$ the potential changes from a minimum at $\phi=0$ to a local maximum at this point (for better contrast, see Fig. \ref{potential}a for $b=1.5$). Then this range of parameters marks the transition from the point where the potential favors the formation of a single kink to that where the potential starts to favor the appearance of two separated kinks. The increasing of $b$ to even larger values (in the direction of the formation of two-kinks) leads to a revival in the number of the two-bounce windows. This makes sense with some known results,
since total suppression of two-bounce windows was not observed in other models of two-kinks \cite{mo,mo2}. This signals that in the model analyzed here the total suppression of two-bounce windows in $K\bar K$ collisions requires beside the existence of more than one vibrational state a hybrid character between a kink and a double kink. Finally we stress that in a sense this work is a counterpart of what presented in Ref. \cite{phi6}, which showed that the formation of two-bounce windows in the $\phi^6$ model even in the absence of an internal vibrational state. That is, this work also poses some limits on the applicability of the CSW mechanism to describe $K\bar K$ collisions.

%%%%%%%%%%%%%%%%%%%%%%%%%%%%%%%%%%%%%%%%%%%%%%%%%%%%%%%%%%%%%%%%%%%%%%%%%%%%%%%%%%%%%%%%%%%%%%%%%%%%%%%%%%%%%%%%%%%%%%%%%%%%%%%%%%%%%

\section{Acknowledgements}
The authors thank FAPEMA, CAPES and CNPq for financial support. Gomes thanks D. Bazeia and R. Casana for discussions.

%%%%%%%%%%%%%%%%%%%%%%%%%%%%%%%%%%%%%%%%%%%%%%%%%%%%%%%%%%%%%%%%%%%%%%%%%%%%%%%%%%%%%%%%%%%%%%%%%%%%%%%%%%%%%%%%%%%%%%%%%%%%%%%%%%%%%


\begin{thebibliography}{99}

\bibitem{dauxois} T. Dauxois, M. Peyrard, {\it Physics of Solitons}, Cambridge University Press, Cambridge (2006).

\bibitem{top} N. Manton, P. Sutcliffe, {\it Topological Solitons}, Cambridge University Press, Cambridge, (2004).

\bibitem{mod}  P. P. Avelino, J. P. M. Carvalho, C. J. A. P. Martins, J. C. R. E. Oliveira, {\it Can our universe be inhomogeneous on large sub-horizon scales?}, Phys. Lett. B 512, 148 (2001).

\bibitem{zko} Ya. B. Zel'dovich, I. Yu. Kobzarev, L. B. Okun', {\it Cosmological consequences of a spontaneous breakdown of a discrete symmetry}, Sov. Phys. JETP 40 1 (1975).

\bibitem{vac} S. W. Hawking, I. G. Moss, J. M. Stewart, {\it Bubble collisions in the very early universe}, Phys. Rev. D 26, 2681 (1982).

\bibitem{brane} M. Bucher, {\it A brane world universe from colliding bubbles}, Phys. Lett. B 530, 181 (2002).

\bibitem{bubble} J. Branden, J. R. Bond, L. Mersini-Houghton, {\it Cosmic bubble and domain wall instabilities I: parametric amplification of linear fluctuations}, JCAP 03 (2015) 007.

\bibitem{k} A. E. Kudryavtsev, {\it Solitonlike solutions for a Higgs scalar field}, JETP Lett. 22, 82 (1975).

\bibitem{s} T. Sugiyama, {\it Kink-Antikink Collisions in the Two-Dimensional $\phi^4$ Model}, Prog. Theor. Phys. 61, 1550 (1979).

\bibitem{m} M. Moshir, {\it Soliton-antisoliton scattering and capture in gamma-phi-4 theory}, Nucl. Phys. B 185 318 (1981).

\bibitem{w1} C. A. Wingate, {\it Numerical Search for a $\phi ^4 $ Breather Mode}, SIAM J. Appl. Math. 43(1), 120-140 (1983).

\bibitem{bk} T. I. Belova, A. E. Kudryavtsev, {\it Quasi-periodic orbits in the scalar classical $\lambda \phi^4$ field theory}, Physica D 32, 18 (1988).

\bibitem{csw} D. K. Campbell, J. S. Schonfeld, C. A. Wingate, {\it Resonance Structure in Kink-antikink interactions in $\phi^4$ theory}, Physica. D 9, 1 (1983).

\bibitem{cps} D. K. Campbell, M. Peyrard, P. Soldano, {\it Kink-antikink interactions in the double sine-gordon equation}, Physica D 19, 165 (1986).

\bibitem{cp} D. K. Campbell, M. Peyrard, {\it Solitary wave collisions revisited}, Phys. D, 18 (1986), pp. 47–53.

\bibitem{aom} P. Anninos, S. Oliveira, R. A. Matzner, {\it Fractal structure in the scalar $\lambda(\phi^2-1)^2$ theory}, Phys. Rev. D  44, 1147 (1991).

\bibitem{gk} V. A. Gani, A. E. Kudryavtsev, {\it Kink-antikink interactions in the double sine-Gordon equation and the problem of resonance frequencies}, Phys. Rev. E 60, 3305-3309 (1999).

\bibitem{gh} R. H. Goodman, R. Haberman, {\it Kink-Antikink Collisions in the $\phi^4$ Equation: The n-Bounce Resonance and the Separatrix Map}, SIAM J. Appl.  Dyn. Syst. 4, 1195 (2005).

\bibitem{gll} V. A. Gani, V. Lensky, M. A. Lizunova, {\it Kink excitation spectra in the $(1+1)$-dimensional $\phi^8$ model}, JHEP 08 (2015)147.

\bibitem{a} S. Aubry, {\it Unified approach to interpretation of displacive and order-disorder systems. II. Displacive systems}, J. Chem. Phys. 64, 3392 (1976).

\bibitem{fkv} Z. Fei, Y. S. Kivshar, L. V\'azquez, {\it Resonant kink-impurity interactions in the $\phi^4$ model}, Phys. Rev. A 46, 5214 (1992).

\bibitem{fkv2} Z. Fei, Y. S. Kivshar, L. V\'azquez, {\it Resonant kink-impurity interactions in the sine-Gordon model}, Phys. Rev. A 45,6019 (1992).

\bibitem{fkv3} Y. S. Kivshar, Z. Fei, L. V\'azquez, {\it Resonant soliton-impurity interactions}, Phys. Rev. Lett., 67
 1177 (1991).

 \bibitem{gh2} R. H. Goodman, R. Haberman, {\it Interaction of sine-Gordon kinks with defects: The two-bounce
resonance}, Phys. D 195, 303 (2004).

\bibitem{ty1} J. Yang, Y. Tan, {\it Fractal structure in the collision of vector solitons}, Phys. Rev. Lett. 85, 3624 (2000).

\bibitem{ty2} Y. Tan, J. Yang, {\it Complexity and regularity of vector-soliton collisions}, Phys. Rev. E 64, 056616 (2001).

\bibitem{gh3} R. H. Goodman, R. Haberman, {\it Vector soliton interactions in birefringent optical fibers}, Phys. Rev.
E 71, 055065 (2005).

\bibitem{bk2} T. I. Belova, A. E. Kudryavtsev, {\it Solitons and their interactions in classical field theory}, Physics-Uspekhi 40, 359 (1997).

\bibitem{phi6} P. Dorey, K. Mersh, T. Romanczukiewicz, Y. Shnir, {\it Kink-Antikink Collisions in the $\phi^6$ Model}, Phys. Rev. Lett. 107, 091602 (2011).

\bibitem{gkl} V. A. Gani, A. E. Kudryavtsev, M. A. Lizunova, {\it Kink interactions in the $(1+1)-$dimensional $\phi^6$ model}, Phys. Rev. D 89, 125009 (2014).

\bibitem{dutra} A. de Souza Dutra, {\it Continuously deformable topological structure}, Physica D 238, 798 (2009).

\bibitem{bmm} D. Bazeia, J. Menezes, R. Menezes, {\it New Global Defect Structures}, Phys. Rev. Lett. 91, 241601 (2003).

\bibitem{bps1} E. B. Bogomolny, {\it The stability of classical solutions}, Sov. J. Nucl. Phys. 24, 449 (1976).

\bibitem{bps2} M. K. Prasad, C. M. Sommerfield, {\it Exact Classical Solution for the $'$t Hooft Monopole and the Julia-Zee Dyon}, Phys. Rev. Lett. 35, 760 (1975).

\bibitem{col1} J-L Gervais, B. Sakita, {\it Extended particles in quantum field theories}, Phys. Rev. D 11, 2943 (1975).

\bibitem{col2} J-L Gervais, A. Jevicki, B. Sakita, {\it Perturbation expansion around extended-particle states in quantum field theory}, Phys. Rev. D 12, 1038 (1975).

\bibitem{col3} N. H. Christ, T. D. Lee, {\it Quantum expansion of soliton solutions}, Phys. Rev. D 12, 1606 (1975).

\bibitem{mo} T.S. Mendon\c ca, H.P. de Oliveira, {\it A note about a new class of two-kinks}, JHEP (06)(2015)133.

\bibitem{mo2}T.S. Mendon\c ca, H.P. de Oliveira, {\it The collision of two-kinks defects}, JHEP (09)(2015)120.


\end{thebibliography}
\end{document}